\documentclass[fleqn,11pt]{article}

\usepackage{amsfonts,amsmath,amssymb}
\usepackage{latexsym,color,cite}

\usepackage[utf8]{inputenc}
\usepackage{graphicx}

\usepackage{amsfonts,amssymb,cite}
\usepackage{graphicx}

\def\onecol{\onecolumn \mathindent 2em}

\def\noi{\noindent}

\newcommand{\Title}[1]{\noi {{\Large\bf #1}}\\[1ex]}

\def\Aunames#1{\noi{\bf #1}}
\def\au#1{${}^{#1}$}
\def\Addresses#1{\medskip\noi \protect
	\begin{description}\itemsep -3pt {\it #1} \end{description}}
\def\adr#1#2{\item[${}^{#1}$]{\it #2}}

\newcommand{\Abstract}[1]{\vskip 2mm \begin{center}
        \parbox{16.4cm}{\small\noi #1} \end{center}\medskip}

\def\email#1#2{\footnotetext[#1]{e-mail: #2}\addtocounter{footnote}{1}}

\def\nqq{\hspace*{-2em}}

\def\qq{\qquad}

\def\inch{\hspace*{1in}}

\usepackage{color}

\def\Acknow#1{\subsection*{Acknowledgments} #1}

\def\Jl#1#2{#1 {\bf #2},\ }

\def\ApJ#1 {\Jl{Astroph. J.}{#1}}
\def\CQG#1 {\Jl{Class. Quantum Grav.}{#1}}
\def\DAN#1 {\Jl{Dokl. AN SSSR}{#1}}
\def\GC#1 {\Jl{Grav. Cosmol.}{#1}}
\def\GRG#1 {\Jl{Gen. Rel. Grav.}{#1}}
\def\IJMPD#1 {\Jl{Int. J. Mod. Phys. D}{#1}}
\def\JETF#1 {\Jl{Zh. Eksp. Teor. Fiz.}{#1}}
\def\JETP#1 {\Jl{Sov. Phys. JETP}{#1}}
\def\JHEP#1 {\Jl{JHEP}{#1}}
\def\JMP#1 {\Jl{J. Math. Phys.}{#1}}
\def\NPB#1 {\Jl{Nucl. Phys. B}{#1}}
\def\NP#1 {\Jl{Nucl. Phys.}{#1}}
\def\PLA#1 {\Jl{Phys. Lett. A}{#1}}
\def\PLB#1 {\Jl{Phys. Lett. B}{#1}}
\def\PRD#1 {\Jl{Phys. Rev. D}{#1}}
\def\PRL#1 {\Jl{Phys. Rev. Lett.}{#1}}

\def\lal{&&\nqq {}}
\def\eq{Eq.\,}
\def\eqs{Eqs.\,}
\def\beq{\begin{equation}}
\def\eeq{\end{equation}}
\def\besub{\begin{subequations}}
\def\esub{\end{subequations}}
\def\bear{\begin{eqnarray}}
\def\bearr{\begin{eqnarray} \lal}
\def\ear{\end{eqnarray}}
\def\earn{\nonumber \end{eqnarray}}

\def\nnn{\nonumber\\ \lal }

\def\yy{\\[5pt] {}}
\def\yyy{\\[5pt] \lal }

\def\dst{\displaystyle}
\def\tst{\textstyle}
\def\fracd#1#2{{\dst\frac{#1}{#2}}}
\def\fract#1#2{{\tst\frac{#1}{#2}}}
\def\Half{{\fracd{1}{2}}}
\def\half{{\fract{1}{2}}}

\def\e{{\,\rm e}}
\def\D{\partial}

\def\im{\mathop{\rm Im}\nolimits}

\def\diag{\mathop{\rm diag}\nolimits}

\def\const{{\rm const}}

\def\then{\ \Rightarrow\ }
\newcommand{\toas}{\mathop {\ \longrightarrow\ }\limits }

\def\eqn#1{\eq\eqref{#1}}
\def\rf{\eqref}

\def\mN{_\mu^\nu}

\def\R{{\mathbb R}}

\def\cF{{\mathcal F}}

\def\tT{{\tilde T}{}}
\def\Veff{V_{\rm eff}}
\def\GR{general relativity}
\def\sph{spherically symmetric}
\def\ssph{static, spherically symmetric}
\def\bh{black hole}
\def\bhs{black holes}
\def\wh{wormhole}

\def\asflat{asymptotically flat}
\def\emag{electromagnetic}
\def\mult{multidimensional}
\def\pb{perturbation}
\def\pbs{perturbations}
\def\Scw{Schwarz\-schild}
\def\Schr{Schr\"odinger}
\def\RN{Reiss\-ner-Nord\-str\"om}
\def\da{\delta\alpha}
\def\db{\delta\beta}
\def\dg{\delta\gamma}
\def\dxi{\delta\xi}
\def\dr{\Delta r}

\begin{document}
\onecol
\thispagestyle{empty}

\Title{5D black holes and mirror (topological) stars from nonlinear\yy electrodynamics: 
		Existence and stability}

\Aunames{Kirill A. Bronnikov,\au{a,b,c,1} Sergei V. Bolokhov,\au{b,2} and Milena V. Skvortsova\au{b,3}}
	
\Addresses{\small
\adr a	{Center of Gravitation and Fundamental Metrology, Rostest, 
		Ozyornaya ul. 46, Moscow 119361, Russia}
\adr b	{Institute of Gravitation and Cosmology, RUDN University, 
		ul. Miklukho-Maklaya 6, Moscow 117198, Russia}
\adr c 	{National Research Nuclear University ``MEPhI'', 
		Kashirskoe sh. 31, Moscow 115409, Russia}
		}
		

\Abstract
  { We consider static, spherically symmetric solutions of 5D general relativity with magnetic fields 
   governed by nonlinear electrodynamics (NED) with the Lagrangian $L(\cF)$, $\cF = F_{AB} F^{AB}$, 
   and show that generic solutions describe either 5D black holes (also called black strings due to a circular
   extra dimension) or so-called mirror stars with perfectly reflecting boundary surfaces (also called 
   topological stars),
   not to be confused with 4D configurations of mirror matter considered in particle physics. 
   Two particular examples of such solutions have been obtained, admitting analytic
   expressions for the metric coefficients and $L(\cF)$, and their stability under radial (monopole)
   perturbations is studied. While the whole obtained family of black hole solutions turns out to be 
   stable, mirror star solutions prove to be stable only in a certain range in the parameter space. 
   We thus extend to the Einstein-NED system the results previously obtained for Einstein-Maxwell fields.
}

\email 1 {kb20@yandex.ru}  
\email 2 {boloh@rambler.ru}
\email 3 {milenas577@mail.ru}

\section{Introduction}

 The quest for extra dimensions often resides in the subatomic realm of high-energy particle 
  physics. However, we may argue that their most dramatic signatures may be hidden in the
  macroscopic world of compact astrophysical objects. We propose that under extreme conditions, 
  the immense space-time curvature effectively displays hidden degrees of freedom of a fifth 
  dimension in the macroscopic world. While standard 4D general relativity and its 
  extensions lead to the formation of event horizons --- surfaces determining the 
  boundary of any causal influence from the interior \cite{bh1,bh2, bh3,bh4,bh5},
  the presence of a hidden extra dimension allows for a radical alternative \cite{kb95a,kb95b}.
 
  The concept of such hypothetical objects, which we would like to call mirror stars, arises directly 
  from multidimensional black hole solutions through a mutual substitution of the original time 
  coordinate and one of those of compact extra dimensions. This replacement naturally leads 
  to a new valid solution, as the underlying field equations are indifferent to which coordinate 
  is interpreted as time and which as ``extra.'' Thus a surface that once served as an event 
  horizon is transformed to a perfectly reflecting boundary, 
  as probably first noticed in 1995 in \cite{kb95a,kb95b}.
  Later on, the possible existence and plausibility of such horizonless geometries was anticipated
  in the string theory related studies aimed at solving the \bh\ information paradox, which led to
  the notion of ``fuzzballs'' as \bh\ microstate geometries \cite{str1, str2, str3, str4} (see also numerous
  references therein). In particular, the term ``topological stars'' for such geometries seems to have been
  proposed in \cite{str4}. Such objects as solutions to Einstein-Maxwell equations with monopole magnetic
  fields have been considered in detail in \cite{tops1}, where the term ``topological stars'' was justified by
  the existence of ``topological cycles supported by a magnetic flux.'' However, as discussed in 
  \cite{we25, we26}, this kind of geometry can even exist as a modified 5D \Scw\ solution and thus a 
  magnetic flux is not necessary, while the circle in the fifth dimension is a usual Kaluza-Klein construct.
   
  The 4D section of the mirror-star geometry might belong to a \wh\ (see, for example, 
  \eqn {ds-ex1} with $r_b > 2m$), but its would-be throat now becomes a reflecting boundary because 
  the extra-dimensional circle shrinks there to a point without leading to a curvature singularity, 
  and any signal or particle getting there has no other way than to return back. 
  
  We have to mention here that the term ``mirror'' is also used in studies of particle physics for 
  an extension of the Standard Model assuming the existence of ``mirror particles'' and ``mirror matter'' 
  that can form ``mirror stars as dark-sector analogues of regular stars that shine in dark photons''
  \cite{curtin20}.  The subject of the present paper as well as those cited above has nothing to do with
  this direction of particle physics.  
 
  The interest in objects like mirror stars (in our meaning) is further fueled by an active 
  discussion of reflection phenomena in modern astrophysics, including various types of 
  gravitational echoes predicted in black hole and wormhole space-timess \cite{ech1,ech2,ech3,ech4}. 
  Notably, recent theoretical and phenomenological studies (e.g., \cite{25-lim}) have 
  established stringent constraints on reflective compact objects whose surface radius 
  $r_s$ is extremely close to a would-be event horizon $r_h$, such that $r_s = r_h(1+\epsilon)$. 
  Specifically, objects with a compactness parameter $1 + \epsilon < 1 + 10^{-3}$ are now
  largely excluded in current models, which makes a study of stability ranges for theoretically 
  consistent models particularly timely.
  
  Examples of 5D mirror star solutions to Einstein-Maxwell equations with monopole magnetic
  fields have been considered in sufficient detail in \cite{tops1, tops2, tops3, we25, we26} 
  along with their \bh\ counterparts, all of them being special cases of more general \mult\ 
  solutions obtained earlier \cite{kb95a,kb95b,bh5}.  In particular, in \cite{tops3}, stability 
  of these mirror star models under nonradial \pbs\ was established, while their stability 
  ranges with respect to radial (monopole) \pbs\ were found in \cite{we26}. The present paper 
  is devoted to a similar study of 5D models sourced by nonlinear electrodynamics (NED),
  being motivated by the fact that extremely strong fields inside and near compact objects
  inevitably trigger self-interaction effects that can be accounted for by introducing nonlinear
  \emag\ field Lagrangians. 
  We demonstrate the generic existence of mirror star and black hole solutions to the 
  5D NED-Einstein equations and obtain some examples of such solutions. Furthermore, we 
  consider their monopole \pbs\ emerging in the system owing to the effective
  scalar nature of the extra-dimensional metric component $g_{55}$. Such \pbs, if admitted
  by a local stationary system, are known to be most likely to cause its instability due to 
  absence of a centrifugal barrier in the corresponding effective potentials, see, e.g., 
  \cite{br-book, kb-hod, bfz11, bkz12}.  
  We prove that the obtained black hole solutions are stable under such \pbs, while mirror 
  star solutions are stable only in a certain range in their parameter space.

\section{General observations}   
\subsection{5D equations and a solution algorithm}

  We will deal with 5D general relativity (GR) with a nonlinear \emag\ field as a source, so the action is 
\beq
			S = \frac 12 \int \sqrt{^5g}\, d^5x \big[R_5 - L(\cF)\big],			
\eeq  
  where $R_5$ is the 5D Ricci scalar, $^5g$ is the determinant of the 5D metric $g_{AB}$ with the
  signature $(+----)$, $L(\cF)$ is a function of the \emag\ invariant $\cF = F_{AB} F^{AB}$, and we are 
  using units in which $8\pi G = c =1$ in standard notations. The metric is assumed in the \ssph\ form
\beq           \label{ds5}
		ds_5^2 = g_{AB} dx^A dx^B 
        = \e^{2\gamma(u)} dt^2 - \e^{2\alpha(u)} du^2 - \e^{2\beta(u)} d\Omega^2 - \e^{2\xi(u)} dv^2,
\eeq    
  where $u$ is the radial coordinate (admitting an arbitrary parametrization), $v$ is the fifth
  coordinate specified on a circle of radius $\ell$, so that $v \in [0, 2\pi\ell$);  
  $d\Omega^2 = d\theta^2 + \sin^2\theta d\varphi^2$, and the coordinates $x^A$ are counted 
  by the rule $(t, u, \theta, \varphi, v) = (0,1,2,3,5)$. The metric is supposed to be \asflat, so that 
  at $r \equiv \e^\beta \to \infty$, we require $\gamma \to 0$, $\xi \to 0$, and $\e^{-\alpha}dr/du \to 1$.       
  Nonzero components of $F_{AB}$ compatible with this space-time symmetry correspond 
  to radial electric ($F_{01} = -F_{10}$), radial magnetic  ($F_{23} = - F_{32}$), and so-called 
  quasiscalar ($F_{15} = -F_{51}$) fields.

  In four dimensions, \ssph\ solutions with either electric or magnetic fields are quite easily found
  in terms of the \Scw\ radial coordinate $r \equiv \e^\beta$, both for a prescribed form of 
  $L(\cF)$ and after choosing a desired metric function $A(r)= \e^{2\gamma}$ 
  \cite{N1, N2, N3, N4, N6}; certain difficulties only arise in a search for dyonic solutions with 
  combined electric and magnetic charges (see \cite{N-dyon, kru-dyon} and references therein). 
  Unlike that, in 5D the problem looks much more involved. We will try to obtain some solutions 
  of interest, assuming for certainty and simplicity that there is only a radial magnetic field, so that 
  among $F_{AB}$ only $F_{23} = - F_{32}$ are nonzero. As shown in \cite{we25, we26},
  electric and quasiscalar fields do not yield 5D solutions to the Einstein-Maxwell equations 
  of equal interest with magnetic fields, and thus we also restrict ourselves to magnetic fields in the 
  context of NED.
  
  Thus we are considering the 5D Einstein equations 
  $G_A^B \equiv R_A^B - \half R \delta_A^B = - T_A^B$ with the stress-energy tensor
  corresponding to such a magnetic field:
\beq       \label{SET}  
		T^A_B = \Half \diag\Big(L, L, L- 2\cF L_\cF, L- 2\cF L_\cF, L\Big),
\eeq   
  where $L_\cF = dL/d\cF$. We have $T^0_0 = T^1_1 = T^5_5$, while 
  the invariant $\cF$ has the same form as in 4D: 
\beq       \label{cF}
		\cF = 2 F_{23} F^{23} = \frac{2 q^2}{r^4(u)} \equiv 2 q^2 \e^{-4\beta},     
\eeq
  where $q$ is the magnetic charge, and $r = \e^\beta$ is the spherical radius.
  
  Let us present the nonzero components of the 5D Ricci tensor for the metric \rf{ds5} 
  without fixing the radial coordinate $u$ (the prime denotes $d/du$): 
\bearr            \label{Ric-gen}
		R^0_0 = -\e^{-2\alpha}\big[\gamma'' +\gamma'(2\beta'+\gamma'+\xi'-\alpha')\big],
\nnn		
  		R^1_1 = - \e^{-2\alpha}\big[2\beta'' + \gamma'' + \xi'' 
    			+ 2\beta'^2 + \gamma'^2 + \xi'^2  
    			 - \alpha'(2\beta' + \gamma' + \xi')\big],
\nnn
  		R^2_2 = R^3_3 = \e^{-2\beta}- \e^{-2\alpha}\big[\beta'' 
  		+ \beta'(2\beta'+\gamma'+\xi'-\alpha') \big],
\nnn
		R^5_5 = - \e^{-2\alpha}\big[\xi'' + \xi'(2\beta'+\gamma'+\xi'-\alpha') \big],	
\ear  
  Note that putting $\xi = \const$, we obtain the Ricci tensor components $R\mN$ of
  the 4D section of \rf{ds5}.  It is also useful to present the Einstein tensor component 
  $G^1_1 = R^1_1 - \half R$ since the corresponding Einstein equation is first-order and 
  is an integral of the others which are second-order:
\bearr            \label{G11}
		G^1_1 = - \e^{-2\beta} + \e^{-2\alpha}
				\big(\beta'^2 + 2\beta'\gamma' 
				+ 2\beta'\xi'  + \gamma'\xi'\big).
\ear 

  As follows from \rf{Ric-gen}, the relation $R^0_0 = R^5_5$ (a consequence of 
  \rf{SET}) is easily integrated giving
\beq            \label{int0-5}
		\xi'-\gamma' = N \e^{-(2\beta + \gamma + \xi- \alpha)}, \qq N = \const.
\eeq    
  Furthermore, the relations $R^1_1 = R^5_5$ and $R^1_1 = R^0_0$ yield, respectively,
\bearr                \label{1-5} 
	2\beta'' + 2\beta'^2 + \gamma'' + \gamma'^2 - (\alpha'+\xi') (2\beta' + \gamma') =0,
\\ \lal              \label{1-0}
	2\beta'' + 2\beta'^2 + \xi'' + \xi'^2 - (\alpha'+\gamma') (2\beta' + \xi') =0.
\ear
  Quite evidently, among the three equations \rf{int0-5}, \rf{1-5}, \rf{1-0} only two 
  are independent.  
  
  As already remarked, it is hard to obtain solutions to the 5D NED-Einstein equations
  with specified $L(\cF)$. However, solutions can be found by specifying one of the metric
  functions, either $\gamma(u)$ or $\xi(u)$. We so far did not fix a choice of the radial 
  coordinate $u$, but after its proper choice, with known $\gamma(u)$ or $\xi(u)$, it becomes 
  possible to determine the whole metric tensor $g_{AB}$ solely from \eqs \rf{int0-5}, 
  \rf{1-5}, \rf{1-0} that follow from symmetry of the tensor \rf{SET}, after that the 
  Lagrangian function $L(\cF)$ is found from the equation $G^1_1 = -T^1_1 = - L(\cF)/2$ 
  in terms of $u$, and its determination in terms of $\cF$ is then possible using \eqn{cF}.
      
  Thus, if we choose $u \equiv r = \e^\beta$ as the radial coordinate, then in \rf{1-5}
  and \rf{1-0} we have $\beta'' + \beta'^2=0$, and with any specified $\gamma(r)$ 
  \eqn{1-5} gives at once an expression for $\alpha'+\xi'$, 
\beq     \label{a-xi}
		\alpha'+\xi' = \frac{\gamma'' + \gamma'{}^2}{2/r + \gamma'},
\eeq
  hence by quadrature we find $\alpha+\xi$. On the other hand, \eqn{int0-5} can be rewritten 
  in the form
\beq            \label{xi-ga}
		2 (\xi' - \gamma')\e^{2\xi - 2\gamma} = \frac{2N}{r^2} \e^{\alpha + \xi - 3\gamma}.
\eeq    
  Integrating it, we learn $\xi - \gamma$, hence $\xi(r)$ (since $\gamma(r)$ was specified), and 
  also find $\alpha(r)$ since $\alpha + \xi$ was previously found. Thus we know the whole metric. 
  
  Our set of equations is symmetric under the replacement $\gamma \leftrightarrow \xi$, hence we 
  can similarly specify $\xi(r)$, use \eqn{1-0} instead of \rf{1-5} to find $\alpha+\gamma$, and
  integrating \rf{int0-5} that takes the form
\beq            \label{ga-xi}
		2 (\gamma'-\xi')\e^{2\gamma - 2\xi} = -\frac{2N}{r^2} \e^{\alpha + \gamma - 3\xi},
\eeq    
  determine $\gamma(r)$ and hence the whole metric. 
  
  Other solutions to the NED-Einstein equations may be found using other coordinate 
  conditions. For example, choosing the coordinate $u$ by putting $\alpha = - \xi + n\gamma$ 
  ($n = \const$) and specifying $\gamma(u)$ ``by hand,'' we make \eqn{1-5} a linear first-order 
  equation for $\beta'(u)$; solving it, we learn $\beta(u)$, then determine $\xi(u)$ using
  \eqn{int0-5}, and lastly find $\alpha(u)$ from our coordinate condition.      
      
\subsection{Possible black hole and mirror star solutions}    

  Before addressing particular examples, let us show in a general form that our
  problem setting allows for obtaining both black hole and mirror star solutions.
  Let us notice that a {\bf black hole} with the metric \rf{ds5} is characterized by 
  the existence of a horizon at some $u = u_h$ at which $g_{tt} = \e^{2\gamma(u)}$ 
  has a regular zero while the spherical radius $r = \e^\beta$ and the 
  extra-dimensional component $g_{55}= -\e^{2\xi}$ remain finite and regular at and 
  outside the sphere $u = u_h$. Similarly, a {\bf mirror star} with the metric \rf{ds5} 
  contains a boundary surface $u = u_b$ at which $g_{vv} = \e^{2\xi} =0$, 
  while the radius $r = \e^\beta$ and the temporal component 
  $g_{00}= \e^{2\gamma}$ remain finite and regular at and outside the sphere $u = u_b$. 
  
  Let us again choose $r$ as a radial coordinate and consider \eqn{xi-ga}.
  Assuming at spatial infinity ($r\to\infty$) $\e^\gamma = \e^\xi = \e^\alpha =1$, the 
  solution of \rf{xi-ga} may be written as  
\beq        \label{xi} 
	\e^{2\xi} = \e^{2\gamma} 
		\bigg(1 - 2N \int_r^\infty \frac{\e^{\xi+\alpha - 3\gamma}}{r^2} dr\bigg).
\eeq    
  The integral converges at $r=\infty$. On the other hand, assuming $2/r +\gamma' >0$ 
  (as is generically the case) and a regular function $\gamma(r)$, \eqn{a-xi} gives a finite
  value of $\alpha'+\xi'$, hence $\alpha+\xi$ is also finite, therefore the integrand in \rf{xi} 
  is finite. All that is true at least at sufficiently large $r$ due to the asymptotic behavior of 
  all functions. Thus an integral from any such $r = r_0$ to infinity is finite, and we can 
  choose the integration constant $N > 0$ in such a way that the quantity $\e^{2\xi}$ will 
  turn to zero at this $r_0$ --- it will mean that we have obtained a mirror star solution where 
  $r = r_0$ is a mirror surface, to be denoted $r=r_b$. The reflecting properties of such 
  surfaces for both massless and massive (but not too massive) particles were proved 
  in \cite{kb95a,kb95b,we25,we26} under the assumption that the extra dimension 
  is sufficiently small to be invisible by modern instruments.   
  
  Thus the system under study admits a large variety of \asflat\ mirror star solutions.
  
  At a mirror surface $r=r_b$, as we saw, generically $\alpha+\xi$ is finite, hence 
  $\e^{\alpha} \sim \e^{-\xi} \to \infty$. Also, since the integrand in \rf{xi} is finite at $r = r_b$, 
  we can assert that $d(\e^{2\xi})/dr$ is also finite there. The basic properties of mirror surfaces 
  are thus observed without specifying the choice of NED, and these properties are in principle
  the same as were previously found with Maxwell electrodynamics. 
  In particular, by full analogy with their behavior at \bh\ horizons, we can assert that all 
  curvature invariants are finite in the limit $r \to r_b$.  Their finiteness, however, does not guarantee 
  regularity of the very point $r=r_b$ which is actually a center of symmetry in the $(r,v)$ subspace
  (see a more detailed description in \cite{we26}). To avoid a conical singularity at $r=r_b$, one 
  has to require local flatness (a correct circumference to radius ratio for small circles), 
  which relates the solution parameters to the compactification length $\ell$ of the extra 
  dimension. Generically this length must be of the same order as $r_b$ itself, as can be seen further 
  in Example 2.
  
  Under the same conditions at infinity, we can use \eqn{ga-xi} and write its solution in the form
\beq        \label{ga} 
	\e^{2\gamma} = \e^{2\xi} 
		\Big(1 + 2N \int_r^\infty \frac{\e^{\alpha + \gamma - 3\xi}}{r^2} dr\Big),
\eeq    
  and consider any value of $r_0$ of $r$ such that the integrand in \rf{ga} and $\e^\xi$ are 
  finite at $r \geq r_0$. Then, choosing some $N < 0$, we get $\e^{2\gamma(r_0)} =0$
  while $\e^{2\gamma} >0$ at $r > r_0$, which means that $r = r_0 = r_h$ is a horizon, and 
  we obtain a \bh\ solution.  Quite similarly to mirror star solutions, we can assert that in 
  generic \bh\ ones we have at the horizon $\e^{\alpha} \sim \e^{-\gamma} \to \infty$
  while $d(\e^{2\gamma})/dr$ is finite. The latter circumstance means that we
  are dealing with a simple horizon like \Scw's.
  
  If we choose $\e^{2\gamma}\sim (r-r_h)^2$ near $r_h$, as happens at an extremal
  horizon, then a solution may be obtained in the same manner, but we can no longer assert
  that $\e^{\alpha} \sim \e^{-\gamma} \to \infty$ near $r = r_h$. 
  
  One more general observation, actually already following from \eqn{int0-5}, is that $N >0$
  corresponds to mirror stars while $N \leq 0$ to black holes.
  
  We have proved the generic existence of \bh\ and mirror star solutions using the 
  coordinate $u \equiv r$. However, as long as the system under study admits such kinds 
  of geometries, their existence does not depend on the choice of coordinates, 
  used only as a tool in our reasoning.  
  
 \section{Examples of static solutions}

\noi {\bf Example 1.} 
   Let us show how our scheme reproduces the known solution 
   \cite{we25,we26,tops1} to the 5D Einstein-Maxwell equations, i.e., with $L(\cF) = \cF$.
   We choose the coordinate $r$ and the function $\e^{2\gamma}$ in the \Scw\ form:
\beq         \label{Scw1}
		u \equiv r = \e^\beta, \qq    \e^{2\gamma} = 1-2m/r,\qq  m = \const >0,
\eeq
   thus $m$ has the meaning of a \Scw\ mass.  Then \eqn{1-5} takes the easily integrable form
\beq      
		\alpha' + \xi' = - \frac{m}{r (r -2m)}  \ \then \ 
		\alpha + \xi = - \Half \ln \frac{r-2m}{r} = -\gamma,
\eeq   
   in accord with the requirement $\alpha(\infty) =  \gamma(\infty) =0$.
   Now we can use \eqn{int0-5} to find $\xi(r)$: we have 
\beq          \label{xi1}
		\xi' -\gamma' = N \e^{-2\beta -2\xi-2\gamma} \ \then \ 
		2(\xi' -\gamma') \e^{2\xi - 2\gamma} 
				= 2 N \e^{-2\beta-4\gamma} = \frac{2N}{(r - 2m)^2},
\eeq   
   whence $\e^{2\xi - 2\gamma} =C_1 - 2N/(r-2m), \ C_1 = \const$.   
   The condition $\xi(\infty) = \gamma(\infty) =0$ leads to $C_1 = 1$, and using the notation 
   $r_b = 2m + 2N$, we obtain $\e^{2\xi(r)} = 1-r_b/r$ and finally arrive at the 5D metric
\beq          \label{ds-ex1}
		ds^2 = \Big(1 -\frac{2m}{r}\Big)\,dt^2 - \frac{r^2 dr^2}{(r-2m)(r-r_b)} -r^2 d\Omega^2
		  		- \Big(1- \frac {r_b}{r}\Big)\, dv^2,
\eeq  
   which describes 5D \bhs\ if $r_b \leq 2m$ and mirror stars if $r_b > 2m$, coinciding with the
   Einstein-Maxwell ones considered in \cite{we25,we26,tops1,tops2,tops3}. 

   The last step in the derivation of this solution is to find $L(\cF) = -2 G^1_1$. We obtain 
   $L = 3m r_b/r^4$, which coincides with $L = \cF$ if we identify $2q^2 = 3m r_b$.   
   The case $r_b =0$ corresponds to a \Scw\ \bh\ plus a trivial extra dimension, 
   while with $m =0$ we obtain the simplest massless vacuum mirror star solution 
   discussed in \cite{we25, we26}.
   
\medskip\noi{\bf Example 2: a mirror star solution.} 
  We choose the same radial coordinate $r$ but $\e^{2\gamma}$ in an extreme \RN\ manner:
\beq         \label{Scw2}
		u \equiv r = \e^\beta, \qq    \e^{2\gamma} =( 1-m/r)^2,
\eeq
   where $m=\const$ is again the \Scw\ mass.  Then \eqn{1-5} gives
\beq         \label{xi2}
		\alpha' + \xi' = - \frac{2m}{r (2r -m)}  \ \then \ 
		\e^{\alpha + \xi} = \frac{4 r^2}{(2r-m)^2}, 
\eeq    
   where the integration constant was chosen so that $\e^{\alpha + \xi} \to 1$ 
   at large $r$. The next step is to use \eqn{int0-5} that gives us now
\beq         \label{xi2a}
		2(\xi' -\gamma') \e^{2\xi - 2\gamma} 
				= 2 N \e^{-2\beta- 3\gamma +\xi + \alpha} 
				= \frac{8N r^3}{(r - m)^3 (2r -m)^2}.
\eeq    
   Integration yields
\beq         \label{I1}
		\e^{2\xi - 2\gamma} = 4N I_1(r) + C_2, \qq
		I_1(r) = \frac 1{2r-m} + \frac{2r-3m}{(r-m)^2} + \frac 6m \log \frac{r-m}{2r-m}.
\eeq   
   The condition $\e^{\xi(\infty)} =1$ leads to $C_2 = 1 - 4N I_1(\infty)$, and 
   we accordingly obtain
\beq         \label{xi2b}
		\e^{2\xi} = \frac{(r-m)^2}{r^2} \bigg[1 + \frac {24N}{m}\ln 2
		+ 4N \bigg(\frac 1{2r-m} + \frac{2r-3m}{(r-m)^2} + \frac 6m \log \frac{r-m}{2r-m}\bigg)
		\bigg].
\eeq   
   Then \eqn{xi2} allows us to obtain the last unknown metric coefficient $\e^{2\alpha}$:
\beq
		\e^{2\alpha} = \frac{16 r^4}{(2 r - m)^4} \e^{-2\xi},  
\eeq
   and lastly, using \rf{G11}, we find $L(\cF) = -2 G^1_1$ as a function of $r$:
\bearr   \nqq        \label{L2}
             L(\cF) =  \frac 1 {8 m r^8}
	\bigg[24 N(2 m^2-2 m r- r^2) (m-2 r)^4 \log \frac{r-m}{r-m/2}
		+m \Big(2 m^6-2 m^5 (16 N+9 r)
\nnn	 \nqq
		+3 m^4 r (80 N+21 r)
		-8 m^3 r^2 (81 N+13 r)+8 m^2 r^3 (86 N+9 r)-96 m N r^4-192 N r^5\Big)\bigg].
\ear   
   At large $r$ the function $ L(\cF)$ behaves as
   $L(\cF) = (9m^2 + 6Nm)/r^4$, so this NED theory has a correct Maxwell limit at small 
   $\cF = 2q^2/r^4$, and we should identify  
\beq          \label{mq2}
			9m^2 + 6Nm = 2 q^2\ \  \then  \ \ q^2 > 9m^2/2 
\eeq  
   since $N >0$ in mirror star solutions. Figure 1a shows that $\e^{2\xi}$ turns to zero at
   certain $r > m$, which precisely corresponds to our expectations for mirror star solutions. 
  Thus the range of $r$ is $r \geq r_b > m$, where $r_b$ is the value of $r$ where $\e^{2\xi}=0$,
  indicating a mirror surface. As previously remarked, the absence of a conical singularity at $r=r_b$
  leads to a relationship between the solution parameters and the compactification length $\ell$ 
  of the fifth dimension. For example, if $N =m$, we have $r_b \approx 3.78 m$ and should require 
  $\ell \approx 10.5 m$, while for $N = 0.2 m$ we obtain $r_b \approx 1.925 m$ and 
  $\ell \approx 8.9 m$.      
   
   The previous discussion concerned $N >0$. Considering $N < 0$, we could hope to see
   a valid \bh\ solution, but unfortunately the metric loses analyticity at the horizon $r=m$ 
   due to the presence of $\log (r-m)$ in \eqn{I1}. A curious nonsingular \bh\ solution emerges 
   in the intermediate case $N=0$, with the metric
\beq              \label{ds-N=0}
		ds^2 = \Big(1- \frac mr\Big)^2 dt^2 - \frac{16 r^6 dr^2}{(r-m)^2(2r-m)^4}
					- r^2 d\Omega^2  -  \Big(1- \frac mr\Big)^2 dv^2,
\eeq   
   it contains an extremal horizon at $r = m$, while beyond it, as $r \to m/2$, it exhibits an
   infinitely long ``tube,'' or horn, along which both $g_{00}$ and $g_{55}$ remain finite.   
   The corresponding source Lagrangian function $L(\cF)$ remains nontrivial, as seen from 
   \eqn{L2} with $N=0$.
   
\begin{figure*}    
\centering     
\includegraphics[width=5.5cm]{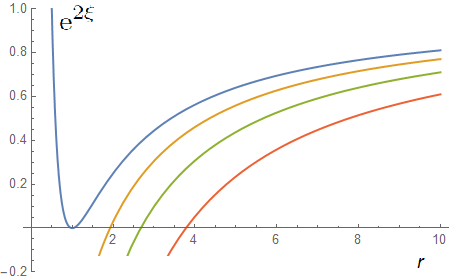}\qq
\includegraphics[width=5.5cm]{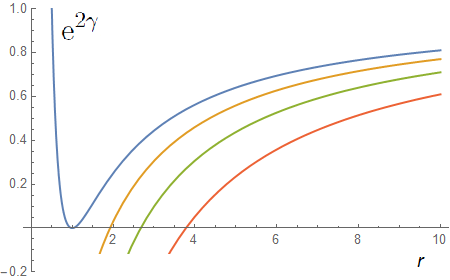}
\caption{\small
		Metric coefficients in Examples 2 (mirror stars) and 3 (black holes). 
		Left: $\e^{2\xi(r)}$ in Example 2, for $m=1$ and $N = 0, 0.2, 0.5, 1$ (upside down). 
		Zeroes of $\e^{2\xi}$ at $N > 0$ correspond to mirror surfaces.
		Right: $\e^{2\gamma(r)}$ in Example 3, for $p=1$ and $N = 0, -0.2, -0.5, -1$ (upside down). 
		Zeroes of $\e^{2\gamma}$ at $N < 0$ correspond to event horizons.
		The curves for $N=0$ in the two panels correspond to the same \bh\ solution with the 
		metric \rf{ds-N=0}, where $\xi \equiv \gamma$.
	}
\end{figure*}         
    
\medskip\noi{\bf Example 3: a \bh\ solution.} 
  Let us use the $\xi \leftrightarrow \gamma$ symmetry of our set of equations and make 
  an assumption similar to \rf{Scw2} but now for $\xi(r)$:
\beq         \label{Scw3}  
  		u \equiv r = \e^\beta, \qq    \e^{2\xi} =( 1- p/r)^2, \qq p = \const >0
\eeq
  (with the notation $p$ instead of $m$ since now it is not a mass). The solution looks the same 
  as in Example 2 with the replacements $\xi \leftrightarrow \gamma$, $m \to p$, and $N\to -N$
  (this integration constant was introduced in \rf{int0-5} in an ``antisymmetric'' manner).
  Thus the two metric coefficients other than \rf{Scw3} can be written as 
\bearr
	\e^{2\gamma} = \frac{(r-p)^2}{r^2} 
			\bigg[ 1 - \frac {24N}{p}\ln 2 - 4N \bigg(\frac 1{2r-p} + \frac{2r-3p}{(r-p)^2} 
					  + \frac 6p \log \frac{r-p}{2r-p} \bigg) \bigg],
\nnn
		\e^{2\alpha} = \frac{16 r^4}{(2 r - p)^4} \e^{-2\gamma}.				  
\ear   
  The same replacements must be made in \eqn{L2} for $L(\cF)$. As already noted,
  \bhs\ are obtained assuming $N < 0$, with the event horizon at $r = r_h > p$ 
  (so that $\e^\xi$ is finite) where $\e^{2\gamma} =0$ (see Fig.\,1b). 
  At large $r$ we obtain $\e^{2\gamma} \approx 1 - 2(p-N)/r$, and
  $L(\cF) \approx 3p (3p- 2N)/r^4$, which leads to identifications for the mass $m$ and 
  magnetic charge $q$:
\beq                      \label{mq3}
		m = p-N, \qq  2q^2 = 3p (3p -2N) \ \ \then \ \ m >p, \qq   q^2 > 9p^2/2.  
\eeq   
   Beyond the horizon (at $r < r_h$) there is a nonstatic region ending at $r =p$ with 
   a singularity due to $g_{55}= \e^{2\xi} \to 0$ while $\e^{2\alpha}$ and $\e^{2\gamma}$ 
   tend to finite negative values: $\e^{2\gamma}  \to 4N/p, \ \e^{2\alpha} \to 4p/N$.  

   At $N =0$ we obtain $m = p$ and again reproduce an extremal \bh\ with the metric 
   \rf{ds-N=0}.
      
   Important requirements to NED theories are those of causality and unitarity, which 
   for $L(\cF)$ theories reduce to the inequalities \cite{usov11, sorokin}
\beq              \label{uni-cau}
		L_\cF \geq 0, \qq    L_{\cF\cF} \leq 0, \qq      L_\cF + 2\cF L_{\cF\cF} \geq 0.
\eeq   
   In our case, $L(\cF)$ is known as a function of $r$ while $\cF = 2q^2/r^4$, and 
   the conditions \rf{uni-cau} reduce to
\beq             \label{uni-cau-r}
		L' \leq 0,  \qq  5L' + r L'' \leq 0, \qq   3L' + r L'' \geq 0.	
\eeq   
    An inspection shows that the function \rf{L2} and the one obtained from it by 
    the replacements $m \to p$, and $N\to -N$, used in Example 3, satisfy the requirements 
    \rf{uni-cau-r} in the relevant ranges of $r$ and the solution parameters. The only exception
    that we could find is a small neighbourhood of the event horizon $r=m$ in the solution 
    \rf{ds-N=0}, where the third inequality \rf{uni-cau-r} is violated.
          
\section{The stability problem}
\subsection{Perturbation equations}   

  Let us consider small time-dependent \sph\ \pbs\ of 5D static solutions with NED 
  like those described above. To do that, we take the metric \rf{ds5} with the functions
  $\alpha, \beta, \gamma, \xi$ depending on both the radial coordinate $u$ and 
  time $t$, only slightly different from their static $u$-dependent expressions, that is, 
\beq                      \label{deltas}
		\alpha(u,t) = \alpha(u) + \da(u,t),  \qq \gamma(u,t) = \gamma(u) + \dg(u,t),
\eeq  
  and similarly for all other quantities, with all ``deltas'' assumed to be small and taken
  into account only in the linear order. 
    
  In 4D \GR\ it is well known that time-dependent spherical \pbs\ of \sph\ metrics cannot 
  be excited by the metric or \emag\ fields but can be induced by \pbs\ of the stress-energy
  tensor of matter, such as scalar fields or continuous matter distributions. It is also
  known that in 5D metrics like \rf{ds5}, when reduced to four dimensions, the component 
  $\e^{2\xi}$, related to possible squeezing or stretching of the extra-dimensional circle, 
  becomes an effective scalar field that admits \sph\ (monopole) \pbs\ able to produce 
  instabilities of the static system (Gregory-Laflamme instabilities \cite{g-laf}). This problem was
  discussed in \cite{we26} for 5D \bhs\ and mirror stars with Maxwell magnetic fields using 
  a 4D form of the theory. Here, unlike \cite{we26}, we study monopole \pbs\ 
  related to $\xi(u,t)$ by directly using the 5D equations.   
  As in previous 4D studies, we separate the \pb\ $\dxi$ from other metric \pbs\
  $\da, \db, \dg$, choosing a particular gauge (that is, a reference frame in perturbed 
  space-time) to simplify the equations. After that it becomes possible to formulate 
  spectral problems whose solutions yield basic information on the \pbs' behavior and
  the system's stability. 

  Let us begin with presenting the nonzero 5D Ricci tensor components for the 
  $(u,t)$-dependent metric \rf{ds5}, neglecting all expressions quadratic in time derivatives:
\bearr            \label{Ric-t}
		R^0_0 = \e^{-2\gamma} (\ddot\alpha +2 \ddot\beta + \ddot\xi)
			-\e^{-2\alpha}\big[\gamma'' +\gamma'(2\beta'+\gamma'+\xi'-\alpha')\big],
\nnn		
  		R^1_1 = \e^{-2\gamma} \ddot\alpha
  				- \e^{-2\alpha}\big[2\beta'' + \gamma'' + \xi'' 
  				+ 2\beta'^2 + \gamma'^2 + \xi'^2
    			 - \alpha'(2\beta' + \gamma' + \xi')\big],
\nnn
  		R^2_2 = R^3_3 = \e^{-2\gamma} \ddot\beta 
		  		+ \e^{-2\beta}- \e^{-2\alpha}\big[\beta'' 
  				+ \beta'(2\beta'+\gamma'+\xi'-\alpha') \big],
\nnn
		R^5_5 = \e^{-2\gamma} \ddot\xi
				- \e^{-2\alpha}\big[\xi'' + \xi'(2\beta'+\gamma'+\xi'-\alpha') \big],	
\nnn
		R_{01} = 2{\dot\beta}' +{\dot\xi}'  - \dot\alpha (2\beta'+\xi') 
					+ 2\dot\beta (\beta'-\gamma') + \dot\xi (\xi'-\gamma'),
\ear    
   where dots and primes denote $\D/\D t$ and $\D/\D u$, respectively. We will use
   the 5D Einstein equations in the form 
   $R^A_B = - \tT^A_B \equiv -(T^A_B - \frac 13 \delta^A_B T^C_C)$.
   
   As to the material source of the metric, we can notice that even with time-dependent 
   metric coefficients the magnetic field has the same form as in the static case,
   so that $\cF = 2q^2 \e^{-4\beta(u,t)}$, and $T^A_B$ is given by \eqn{SET}. In particular, 
   $T_{01} =0$, hence $R_{01} =0$, and 
\beq
		\tT^0_0 =\tT^1_1 = \tT^5_5 = \frac 23(-L+2\cF L_\cF),\qq
		\tT^2_2 =\tT^3_3 = - \frac 23 (L+\cF L_\cF).
\eeq    
   
   All equations can be split into the static ``background'' parts, assumed to be 
   satisfied, and \pb\ parts to be further considered. We will do it choosing the \pb\ gauge
\beq                 \label{gauge}
		2\db + \dxi =0.
\eeq   
   Then the $({}^5_5)$ component of the perturbed 5D Einstein equations can be written as
\beq                 \label{eq-dxi}
           - \e^{2\alpha-2\gamma} \delta\ddot\xi
				+  \dxi'' + \dxi'(2\beta'+\gamma'+\xi'-\alpha') +\xi'(\dg' - \da')
				 = \delta(\tT^5_5 \e^{2\alpha}).	
\eeq   
  It will become a master equation for $\dxi(u,t)$ after expressing the involved metric 
  \pbs\ $\da$ and $\dg$ in terms of $\dxi$, which can be implemented using 
  the Einstein equations $R_{01}=0$ and $2({}^2_2)+({}^5_5)$. Thus, under the condition 
  \rf{gauge}, the equality $R_{01}=0$ reduces to 
\beq
		\dot\alpha (2\beta'+\xi') + \dot\xi (\beta'-\xi') =0,
\eeq  
  which may be directly integrated in $t$ because, due to smallness of time derivatives,
  the quantities $\beta'$ and $\xi'$ should be taken as only $u$-dependent ones.
  Neglecting the arbitrary function of $u$ (the ``integration constant'')
  since only time-dependent \pbs\ are of interest for us, we obtain $\da$ in terms 
  of $\dxi$ and background static quantities:
\beq           \label{da}
		\da = \frac {\xi'-\beta'}{2\beta'+\xi'} \dxi.
\eeq   

  The Einstein equation $2({}^2_2)+({}^5_5)$ allows for expressing one more 
  metric \pb\ involved in \eqn{eq-dxi}, $\dg' - \da'$, in terms of $\dxi, \da$ and the 
  background functions:
\beq
		\dg' - \da' = \frac 1 {2\beta'+\xi'}	\Big\{ \delta(\e^{2\alpha-2\beta}) 
				+ \delta\big[\e^{2\alpha}(2\tT^2_2 + \tT^5_5)\big] \Big\},
\eeq
  where, according to \rf{gauge}, $\db= -\dxi/2$, while the stress-energy tensor 
  components are functions of $\cF = 2q^2 \e^{-4\beta}$, hence their \pbs\ are expressed
  via $\db$, and $\delta\cF = 2\cF \dxi$, and also, $2\tT^2_2 + \tT^5_5 = - L(\cF)$. 
  Substituting all that into \eqn{eq-dxi}, we arrive at the master equation
\bearr                       \label{master}
		- \e^{2\alpha-2\gamma} \delta\ddot\xi
				+  \dxi'' + \dxi'(2\beta'+\gamma'+\xi'-\alpha')  - U(u) \dxi =0,
\yyy                     \label{U}
		U(u) = \e^{2\alpha}\bigg\{
				\frac{2\xi'(\xi'-\beta')}{(2\beta'+\xi')^2}(L -2\e^{-2\beta})
				+ \frac 23 \frac{\xi'-\beta'}{2\beta'+\xi'}(-L + 2\cF L_\cF)
\nnn \inch				
				+ \frac{2\xi'}{2\beta'+\xi'}(\cF L_\cF - \e^{-2\beta})
				+\frac 23 (\cF L_\cF + 2 \cF^2 L_{\cF\cF}) \bigg\},						
\ear
  where $ L_\cF = dL/d\cF$ and $L_{\cF\cF} = d^2 L/d\cF^2$. Since $L$ is determined
  from the equations as a function of $u$ rather than $\cF$, it can be helpful to 
  substitute in \rf{U} 
\beq                 \label{L_FF}
		\cF L_\cF = - \frac{L'}{4 \beta'}, \qq
		\cF^2 L_{\cF\cF} = \frac 1{16}
		\bigg(\frac{4L'}{\beta'}+ \frac{L''}{\beta'{}^2} - \frac{L'\beta''}{\beta'{}^3}\bigg).
\eeq     

   A next step is to reduce \eqn{master} to a canonical form.  
   The radial coordinate $u$ was not yet specified, but a canonical form of the wave
   equation requires a transition to the ``tortoise'' coordinate $z$ 
   (such that $g_{tt}= -g_{zz}$) and a replacement of the unknown function 
   $\dxi$ to get rid of the first-order derivative $\dxi'$:
\beq                 \label{xz}
       du = \e^{\gamma - \alpha} dz, \qq
   	\dxi (u,t) = \e^{-\eta} \psi(u,t), \qq	  \eta := \beta + \xi/2.
\eeq
  As a result, \eqn{master} takes the canonical form $\ddot\psi - \frac {d^2\psi}{dz^2} + \Veff(z) \psi=0$
  with the effective potential 
\beq 			\label{Veff}
		\Veff(z) = \e^{2\gamma - 2\alpha}
				\Big[U + \eta'' + \eta'(\eta'+\gamma'-\alpha')\Big].
\eeq
  Further assuming $\psi(z) = \e^{i\omega t} Y(z)$, where $\omega =\const$ is, in general, 
  a complex frequency, we convert the master equation to the \Schr-like form
\beq  			\label{Schr}
		\frac {d^2 Y}{dz^2} + \big[\omega^2 - \Veff(z)\big] Y=0.
\eeq
  that allows for a further stability analysis. 
   
  We should note here that, despite using the particular gauge \rf{gauge} in the derivation,    
  \eqn{Schr} and the potential $\Veff$ are still gauge-invariant and thus able to
  describe real physical \pbs. This invariance can be directly verified quite similarly to
  \cite{bfz11} by explicitly invoking gauge transformations, and it is also confirmed by the 
  observation that $\Veff$ contains only functions involved in the background static solution.
  One can also notice that \eqs \rf{deltas}--\rf{Schr} are written without fixing the radial
  coordinate $u$ in the background solution. 
  
\subsection{Boundary-value problem: mirror stars}  
  
  Now let us choose, as before, the spherical radius $r$ as the radial coordinate and using it,  
  show that the asymptotic behavior of the potential $\Veff$ at large radii as well as at mirror
  surfaces ($r=r_b$) and \bh\ horizons ($r = r_h$) can be described in a general form even
  without knowledge of particular background or perturbed solutions. We will suppose that 
  at large radii where the \emag\ field is weak, NED turns into its Maxwell limit, and in 
  full analogy with \cite{we26} (where 5D Einstein-Maxwell fields were considered), we have 
  $\Veff \approx \const/r^3$, and $z \approx r$. 
  
  Concerning {\bf mirror surfaces}, let us recall from Sec.\,2.2 that near $r=r_b$ the derivative 
  $d\e^{2\xi}/dr$ is generically finite, consequently, we can write $\e^{2\xi}\sim \dr \equiv r-r_b$. 
  Furthermore, since $\alpha + \xi$ is finite, we have $\e^\alpha \sim 1/\sqrt {\dr}$, and since 
  $\gamma$ is finite, we have
\beq                 \label{z-rb}
		\e^{\alpha-\gamma} \approx \frac k {\sqrt {\dr}}, \ \ \  k = \const
		\ \ \then\   z = \int \e^{\alpha-\gamma} dr \approx 2k \sqrt {\dr} + z_0,
\eeq  
  where let us put $z_0=0$ by choosing the zero point of $z$ at $r=r_b$. On the other hand, 
  in the expression \rf{Veff} it can be easily verified that the term 
  $\e^{2\gamma - 2\alpha} U$ is finite at $r=r_b$ (assuming that $L(\cF)$ 
  is a smooth function), and the behavior of $\Veff$ is determined by the remaining terms, 
  where $\gamma'$ is finite, while in the leading approximation
\beq                 \label{Veff-rb}
		\eta' \approx -\frac{\alpha'}2 \approx \frac 1{4\dr}, \ \ \
		\eta'' \approx -\frac 1{4\dr^2}, \ \ \ 
		\e^{2\gamma - 2\alpha} \approx \frac {\dr}{k^2}   \ \then \
		\Veff \approx  - \frac 1{16 k^2 \dr} \approx - \frac 1{4 z^2},
\eeq  
  where we have used the expression \rf{z-rb} for $z$ near $r=r_b$. Thus the asymptotic behavior 
  of the effective potential of \sph\ \pbs\ of magnetic mirror stars in 5D NED-Einstein theory
  is universal and does not depend on the choice of NED. With this asymptotic behavior of 
  $\Veff$ we can write the approximate solutions to \eqn{Schr} at large radii and $r\to r_b$:
\bearr			\label{Y-inf}
		z\to \infty:\qq      Y(z) = C_1 \e^{i\omega z} + C_2 \e^{-i\omega z},
\\  \lal			\label{Y-rb}
		z\to 0: \qq\ \,      Y(z) = \sqrt{z}(C_3 + C_4 \log z),
\ear       
  where all $C_i$ are integration constants, to be constrained by appropriate boundary conditions.
  
  We are going to study the stability of a system left to itself, without external energy supply,
  and it is natural to require that the total \pb\ energy should be finite. For $\xi$ as an effective
  massless scalar field, this means that at large radii its energy density must be $o(1/r^3)$, 
  which leads to $\dxi' \sim (Y/r)' = o (1/r)$.
  Meanwhile, if $\im\omega \ne 0$, in \eqn{Y-inf} one term exponentially decreases as 
  $z\approx r \to \infty$, while the other exponentially grows, being incompatible 
  with $(Y/r)' = o (1/r)$. Thus only the decreasing term is admissible, and we must 
  simply require $Y \to 0$ as $z \to \infty$.

  At the other boundary $z=0$, the conditions are not so evident. In some previous studies, 
  where the potential $\Veff$ behaved in the same way, the surface $z=0$ was a singularity 
  (see, e.g., \cite{kb-hod, we23}) with a background scalar $\xi \sim \log(r-r_b)\sim \log z$.
  A possible \pb\ $|Y|\sim \sqrt{z} \log z$ corresponds to $\dxi \sim \log z$, blowing up in the 
  same manner as $\xi$ itself, that looked quite admissible. The boundary condition then reads 
  $|Y| \lesssim \sqrt{z} \log z$ and is satisfied by all solutions to \eqn{Schr} with any finite 
  $\omega$, including those which grow in time with any increment $|\im\omega|$. We thus have 
  to infer that our singular object is catastrophically unstable, or, more precisely, its any 
  perturbations immediately turn into a nonlinear regime.
    
  Unlike that, in our present case, the surface $z=0$ ($r=r_b$) is regular in 5D space-time, 
  so it is natural to require that it remains regular under physically admissible \pbs.
  It can be explicitly shown using the Einstein equations that this condition implies 
  $|Y| \lesssim \sqrt{z} \ \then\ C_4 =0$ (see more details in \cite{we26} where this
  problem was discussed for the Einstein-Maxwell equations). 

  We conclude that our stability study requires solving a boundary-value problem for
  \eqn{Schr} with the boundary conditions
\beq           \label{BC-ms-z}
			Y \to 0  \ \ {\rm as}\ \  z\to \infty,
			\qq
			|Y|/\sqrt{z} < \infty  \ \ {\rm as}\ \  z\to 0.
\eeq     

\subsection{Boundary-value problem: black holes}  

   Everything said above on spatial infinity ($z\to \infty$) for mirror star solutions remains 
   valid in the case of \bh\ solutions. The other end of the static region is now the event horizon 
   $r = r_h$ where $\e^{2\gamma} \to 0$, and generically (see Sec. 2.2) the derivative 
   $d\e^{2\gamma}/dr$ is finite, which implies that $\e^{2\gamma} \sim r-r_h$, and 
   $\gamma' \sim 1/(r-r_h)$.  Since both $r = \e^\beta$ and $\xi$ are finite there, and assuming 
   smoothness of $L(\cF)$ (which is in general not guaranteed for all solutions), we can see that 
   the quantity $U$ in \eqn{master} behaves like $\e^{2\alpha} \sim \e^{-2\gamma} \to \infty$ as 
   $r\to r_h$. For the effective potential $\Veff$ we see in \rf{Veff} that the first term, containing $U$, 
   behaves like $\e^{2\gamma} \sim r-r_h$. Furthermore, since $\eta$ and its derivatives are finite near
   $r_h$, while $\gamma' \approx -\alpha' \sim 1/(r-r_h)$, we see that the remaining part of 
   the expression \rf{Veff} is also proportional to $r-r_h$, and it follows that the whole 
   $\Veff \sim r-r_h \to 0$. 
   
   As $dz = \e^{\alpha -\gamma}dr \sim dr/(r-r_h)$, we have $z \sim \log(r-r_h) \to -\infty$ as
   $r\to r_h$: the tortoise coordinate tends to minus infinity, as in other \bh\ solutions. In terms of $z$, 
   $\Veff$ decays exponentially as $r\to r_h$.
   
   All that concerned generic, or simple horizons. If the function $e^\gamma$ is chosen so 
   that the horizon is extremal, that is, $e^{2\gamma}\sim (r-r_h)^{-2}$ (as in our Example 2), then 
   \eqn{a-xi} leads, in general, to a finite value of $\alpha(r_h)$.  (It happens because 
   $\gamma'' + \gamma'^2 =0$ if $\gamma' = 1/(r-r_h)$, while now it is the leading term in $\gamma'$, 
   and so the sum $\alpha'+\xi'$ is determined by subleading terms in $e^\gamma$.) With a finite 
   $\alpha$ and again assuming smooth and finite $L(\cF)$, $U(r_h)$ is also finite, 
   and the leading term in \rf{Veff} is then proportional to $\gamma'\e^{2\gamma} \sim r-r_h$,  that is, 
   vanishes at $r_h$ in the same way as with a simple horizon. We have again 
   $dz \sim dr/(r-r_h) \then z \sim \log(r-r_h) \to -\infty$, and the boundary conditions at
   the horizon look the same for simple and extremal horizons. 
   
   Specifically, the general solution to \eqn{Schr} at $z \to -\infty$ has a form similar to \rf{Y-inf}:   
\beq			\label{Y---inf}
		z\to -\infty:\qq      Y(z) = C_5 \e^{i\omega z} + C_6 \e^{-i\omega z},   
\eeq
  and to exclude \pbs\ that blow up at the horizon, we come again to the condition  $Y \to 0$. 
  Thus for \pbs\ of \bh\ solutions with $\im\omega \ne 0$ we have the boundary conditions 
\beq              \label{BC-bh-z}
                    Y \to 0 \ \ \ {\rm as}\ \ \ z \to \pm\infty.
\eeq    

\subsection{Some features of numerical studies}     

  With the above boundary conditions, the function $Y(z)$ is square-integrable, and it is then 
  straightforward to show that the \Schr\ operator $-d^2/dz^2 + \Veff(z)$ is self-adjoint in the 
  corresponding Hilbert space. Then, according to the Sturm-Liouville theory, its spectrum is 
  purely real, $\omega^2\in \R$. Therefore, in our search for unstable modes of \pbs, we 
  must seek solutions to \eqn{Schr} with $\omega^2 <0$, hence pure imaginary 
  $\omega$, and $Y(z) \sim\e^{-|\omega| z}$ as $z\to \infty$.

  In this approach we study \pbs\ emerging in our system itself, without any signal or energy 
  pumping from outside. This approach is different from the one generally used in 
  studies of quasinormal modes \cite{bh4, qnm09}, where one requires that only outgoing waves 
  are present, and one admits a field $\Phi \sim \e^{-i\omega(t-z)}$ with $\im\omega <0$, so that
  a signal decreases in time but grows at large $z$. Nevertheless, we can notice that
  in our case unstable \pbs\ growing exponentially in time and decaying at infinity
  also correspond to outgoing waves at large $z$. 
    
  Passing on to numerical studies, we face a certain difficulty due to the fact that the effective       
  potential $\Veff$ is actually expressed in terms of an original coordinate $u$, 
  in which the background solution was obtained, and to convert it to the tortoise coordinate $z$ 
  one needs to solve a transcendental equation. Therefore, it looks reasonable to use in 
  practical calculations \eqn{master} expressed in terms of $u$, and further on we will assume 
  $u=r$ since it is the spherical radius $r$ that is used in our examples.  Putting 
  $\dxi(r) = \e^{i\omega t} X(r)$ [so that $X(r) =  \e^{-\eta} Y(z) \equiv (\e^{-\xi/2}/r) Y(z)$], we 
  rewrite \eqn{master} in the form
 \beq   			\label{X''}
		X'' + (2\beta'+\gamma'+\xi'-\alpha') X'
					+\big(\e^{2\alpha-2\gamma}\omega^2 - U(r)\big) X =0,
\eeq   
  where primes denote $d/dr$, and $U(r)$ is given by \eqn{U}.
  
  The boundary conditions for mirror star and \bh\ solutions (\rf{BC-ms-z} and \rf{BC-bh-z},
  respectively)  are equivalently rewritten in terms of $X(r)$ as follows: 
\bearr                          \label{BC-ms-r}
	  \text{Mirror stars:}\qq	X \toas_{r\to \infty} 0, \qq  |X(r_b)| < \infty,
\\ \lal	  		            \label{BC-bh-r}
	  \text{Black holes:}	\ \qq  X \toas_{r\to \infty} 0, \qq X(r_h) =0,  
\ear 
  where we have taken into account that the factor $\e^{-\eta}$ is finite at spatial infinity and 
  at \bh\ horizons but behaves as $\sqrt{z}$ near a mirror surface.
  
  For mirror stars it is important to note that $r=r_b$ is a singular point of \eqn{X''},
  and an attempt to solve it near this point would lead to a numerical instability.
  Therefore, to launch the shooting procedure aimed at a numerical search for eigenvalues
  $\omega^2$, we have to specify initial conditions for $X(r)$ at some regular point close
  to $r_b$. By \rf{BC-ms-r}, an admissible solution must be finite at $r_b$, say, $X(r_b) = X_0$. 
  Suitable conditions at a close point $r_0 = r_b +y$ can be found by asymptotically solving 
  \eqn{X''} near $r = r_b$. To do that, let us estimate the terms in \eqn{X''} at small $y = r-r_b$.
  
  Since $\beta' = 1/r$ and $\gamma'$ are finite at $r=r_b$ while $\e^{2\xi} \sim y$
  and $\alpha'+\xi'$ is finite, we can write $\xi' \approx -\alpha' \approx 1/(2y)$, and the whole 
  factor before $X'$ in \eqn{X''} is $1/y + O(1)$. Furthermore, both $\e^{2\alpha-2\gamma}$ 
  and $U(r)$ behave as $1/y$ due to the factor $\e^{2\alpha}$. Indeed, in \eqn{U} we notice that 
  in each term in the curly brackets the large quantities $\xi'$ and $\xi'^2$ cancel, and the 
  whole expression there reads near $r=r_b$
\[
	U \e^{-2\alpha} \approx \frac 6{r^2} + \frac 43 L(\cF) + 4\cF L_\cF + \frac 43 \cF^2 L_{\cF\cF}.
\]  
  Therefore, \eqn{X''} near $r = r_b$ takes the form
\beq                   \label{X''-rb}
		X'' + \frac{X'}{y} + \frac Ky X =0,  \qq 
			\frac Ky = \Big[\e^{2\alpha-2\gamma}\omega^2 - U(r)\Big]_{r \to r_b},
\eeq  
  so, assuming smoothness of $L(\cF)$, the constant $K$ is found from the background solution 
  evaluated at $r=r_b$.  Now, with $X(r_b) = X_0$, the solution to \eqn{X''-rb} in the leading
  approximation reads $X = X_0 (1 - K y)$, and therefore the complete equation \rf{X''} can be 
  solved numerically under the following initial conditions specified at some $r = r_b + y_0$ 
  with small $y_0$:
\beq                  \label{IC-ms}
		X\big|_{r_b+y_0} = X_0 (1 - Ky_0), \qq   X'\big|_{r_b+y_0} = -K X_0.
\eeq  
  
  For \bh\ solutions, at the horizon $r = r_h$ we have again a singular point of \eqn{X''} and also
  have to move a little aside from this point, say, to $r = r_h+y_0$, to specify the initial conditions.
  The admissible solution for $X(r)$ near $r = r_h$ can now be determined from the one for $Y(z)$ 
  as $z \to -\infty$ since, first, $X (r) =\e^{-\eta} Y(z) \sim Y(z)$, while $\eta = \log r + \xi/2$ is finite 
  at $r=r_h$; second, we know that $z \sim \log(r-r_h)$ near $r_h$, therefore, 
  $\e^z \approx (r-r_h)^b$, where the exponent $b$ is determined by a particular 
  background solution. Therefore, according to \rf{Y---inf}, near $r_h$ an admissible $X(r)$
  behaves as $X \sim (r - r_h)^{b|\omega|}$, and so \eqn{X''} can be solved numerically 
  under the following initial conditions specified at some $r = r_h + y_0$ with small $y_0$: 
\beq                  \label{IC-bh}
		X\big|_{r_h+y_0} = X_0, \qq   X'\big|_{r_h+y_0} = b|\omega| X_0/y_0,
\eeq  
  where $X_0$ is arbitrary since \eqn{X''} is linear.  
  
\section{Examples of stability studies}   
  
\noi {\bf Example 1:} The Einstein-Maxwell solutions characterized by \eqs \rf{Scw1}--\rf{ds-ex1}. 
  According to \cite{we26}, the mirror star solutions with $r_b = 2q^2/(3m) > 2m$ exist at $q^2 > 3m^2$ 
  and prove to be linearly stable under monopole \pbs\ if $r_b < r_b^{\rm crit} \approx 4.004 m$ 
  and are unstable otherwise, while all \bh\ solutions, having horizons at $r_h =2m$ and existing at 
  $q^2 \leq 3m^2$ are stable.
  
\medskip\noi{\bf Example 2.}  
  We analyze the stability of the mirror star solution obtained in Sec.\,3, \eqs \rf{Scw2}--\rf{mq2}. 
  Without loss of generality, we put $m=1$ (fixing the length scale of the problem) and $X_0=1$, thus 
  there remains only one nontrivial free physical parameter $N$. Figure 2 shows the behavior of the effective 
  potential $\Veff$ for \pbs.
\begin{figure}    
\centering     
\includegraphics[width=5.5cm]{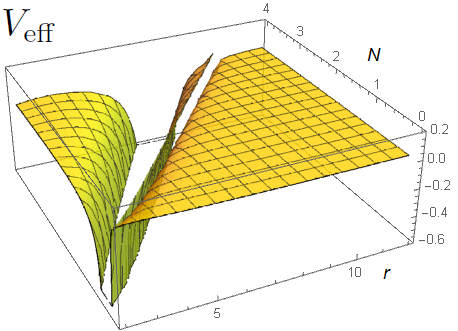}
\caption{\small 
  		The effective potential $\Veff(r)$ for mirror star solutions of Example 2. 
  		Only the surface to the right of the deep holes is relevant.
  		}     \label{Veff-ex2}
\end{figure}
  Since the potential is partly negative, we seek unstable modes among solutions to \eqn{X''}, 
  denoting $X(y) = X(r)|_{r=r_b+y}$. We seek eigenvalues $\omega^2 < 0$ for which the corresponding 
  solution $X(y)$ tends to zero as $y\to \infty$, satisfying the boundary conditions \eqref{IC-ms} near the 
  mirror surface at some small $y_0$. The existence of at least one such negative eigenvalue means that 
  the background solution is unstable. 
   
  We use the so-called shooting method based on the standard Runge-Kutta procedure of solving the 
  Cauchy problem for \eqn{X''} with the above initial conditions in the range $y\in(y_0, y_1)\sim(10^{-3}, 10^3)$ 
  that  yields a numerical accuracy sufficient for our purposes. At that, the value of $\omega^2$ is 
  regarded as a parameter ranging in some interval $({\omega^2_{\min}, 0})$. Each test value of 
  $\omega^2$ determines a numerical solution $X_{\rm num}(y; \omega)$, whose value at the right 
  end $y_1$ is small if $\omega^2$ is close to an eigenvalue of the problem and drastically 
  diverges otherwise. Thus we reveal unstable modes by tracking the behavior of the curve 
  $X_{\rm num}(y; \omega)$ on the right end $y_1$ for various values of $N$.
  \begin{figure}    
  \centering     
  \includegraphics[width=5cm]{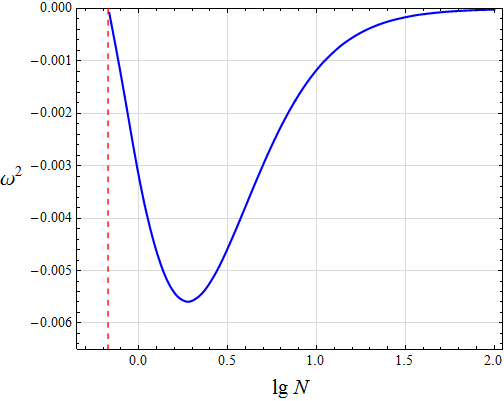}\qq
  \caption{\small 
  		The eigenvalue $\omega^2$ as a function of $\lg N$. The red dashed line corresponds 
  		to $N = N_{\rm crit}\simeq 0.6719642$, showing the left boundary of the instability region.
  		}     \label{omega-n-plot}
  \end{figure}
  \begin{figure}    
  \centering     
  \includegraphics[width=5.5cm]{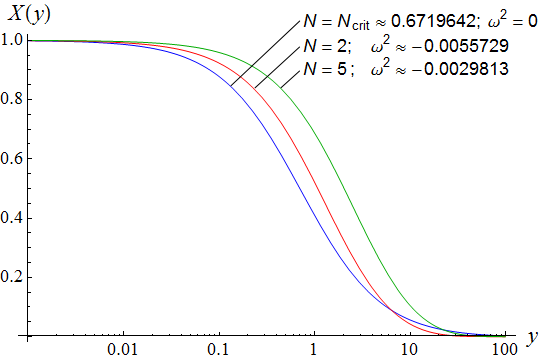}\qq
  \includegraphics[width=5.5cm]{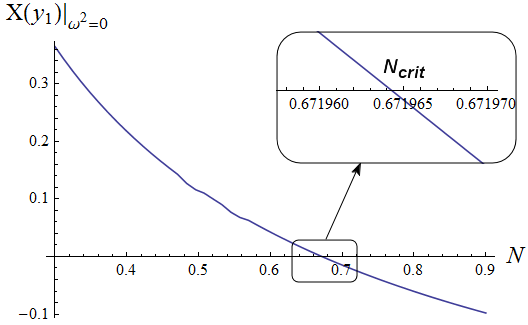}
  \caption{\small 
  		Left: numerical curves $X_{\rm num}(y)$ for various $\omega^2$ and $N$. 
  		Right: The right-end value $X_{\rm num}(y_1)\bigr|_{\omega^2=0}$ as 
  		a function of $N$ allows one to find the critical value $ N_{\rm crit}\simeq 0.6719642$.
  		}     \label{Xnum-plots}
  \end{figure}
\begin{figure*}    
\centering     
\includegraphics[width=6cm]{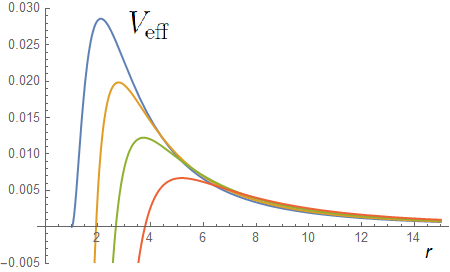}\qq
\includegraphics[width=5.5cm]{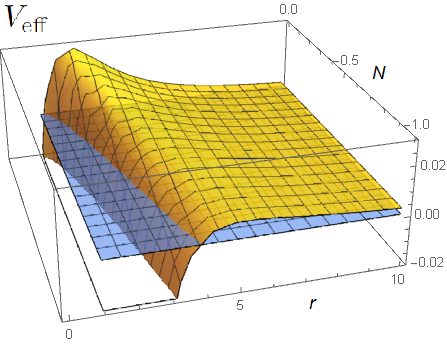}
\caption{\small 
  		The effective potential $\Veff(r)$ for \bh\ solutions of Example 3. Left: $\Veff(r)$ for $N=0, -0.2, -0.5, -1$
  		(upside down). Right: a 3D picture, in which the line $\Veff =0$ corresponds to $r=r_h$, and the 
  		transparent zero level illustrates that $\Veff > 0$ outside the horizon.
  		}     \label{Veff-ex2}
\end{figure*}

  The results are presented in Fig.\,\ref{omega-n-plot}. The plot shows the existence of the negative eigenvalues
  $\omega^2$ revealing the instability region $N > N_{\rm crit} \simeq 0.6719642$ that corresponds to the limiting 
  case $\omega^2=0$. It is calculated numerically from the condition of vanishing of the right-end value 
  $X_{\rm num}(y_1)\bigr|_{\omega^2=0}\to 0$ as $N\to N_{\rm crit}$. At $N < N_{\rm crit}$, eigenvalues
  are absent and thus the static solution is stable. At large $N$ (at least $\sim 1000$) the eigenvalue $\omega^2$ 
  tends asymptotically to zero. This can in principle indicate a possible existence of the right end of the instability 
  region at large $N$ outside the adopted accuracy of numerical analysis. Figure~\ref{Xnum-plots} (left panel) 
  shows examples of numerical curves $X_{\rm num}(y)$ for various $\omega^2$ and $N$ and the method of 
  calculating the critical value $N_{\rm crit}$  (right panel). 
      
\medskip\noi{\bf Example 3.}  For \bh\ solutions given by \eqs \rf{Scw3}--\rf{mq3}, the effective potential is
  shown in Fig.\,\ref{Veff-ex2}. Since $\Veff(r) > 0$ outside the horizon, we conclude that all \bh\ solutions
  in question are stable under monopole \pbs.   

\section{Concluding remarks}   
  
  Considering magnetic \ssph\ solutions for the 5D Einstein-NED system, we have shown that generic
  solutions describe either black holes (also called black strings) or mirror stars (also called topological 
  stars). This substantially extends the results previously obtained for the Einstein-Maxwell system. Our 
  analysis of monopole \pbs\ of such solution reveals a universal behavior of their effective potentials 
  $\Veff$ at small and large radii, independent from a particular choice of the function 
  $L(\cF)$. It, however, turns out to be quite difficult to obtain particular solutions: even in the ``inverse
  problem method,'' when we specify one of the metric functions and only at the end 
  determine the form of $L(\cF)$, we typically come across expressions in quadratures that do not admit 
  a closed analytic form.  
  
  Our two simple examples of the solutions, one for mirror stars and the other for black holes, are 
  qualitatively similar to those of the Einstein-Maxwell system. Thus, by \eqn{mq2}, mirror stars exist at
  $q^2/m^2 > 9/2$, while in the Einstein-Maxwell case at $q^2/m^2 >3$. In both cases, they are stable
  under monopole \pbs\ with sufficiently small charges: roughly, at $q^2/m^2 < 13/2$ and $q^2/m^2 < 6$,
  respectively. And in both cases, all \bh\ solutions turn out to be stable.  
   
  The relevance of a nonlinear description of \emag\ fields is justified by a strong field regime and large 
  space-time curvatures close to \bh\ horizons or mirror surfaces. It would also be of great interest 
  to analyze in such regions QFT effects such as the Casimir effect or particle production, which can be 
  a subject of further studies. 
  
  In any case, from the viewpoint of observations, mirror stars, having no 4D counterparts, if once discovered,
  may be called ``messengers of the fifth dimension.'' Their possible role as one of dark matter  
  candidates can also be kept in mind.

\Acknow{The work of K.B. was partly supported by the Ministry of Science and Higher Education 
     of the Russian Federation, Project FSWU-2026-0010. }


\end{document}